\documentclass[a4paper,11pt]{article}
\usepackage[utf8]{inputenc}
\pdfoutput=1
\usepackage{graphicx}
\usepackage{amsmath,amssymb,mathrsfs}
\usepackage{bbm}
\usepackage{color}
\usepackage{xcolor}
\usepackage{dsfont} 
\usepackage{cancel}
\usepackage{dsfont}
\usepackage{epstopdf}
\usepackage{epsfig}
\usepackage{bm}
\usepackage{authblk}
\usepackage{dcolumn}
\usepackage{enumitem}
\usepackage{multirow}
\usepackage{lineno}
\usepackage{mathtools}
\usepackage{url}
\usepackage{lineno}
\usepackage{wrapfig}
\usepackage{caption}
\usepackage{subcaption}
\usepackage{jheppub}
\usepackage[toc,page]{appendix}
\usepackage{wrapfig}
\usepackage[T1]{fontenc}

\makeatletter 
\gdef\@fpheader{}
\g@addto@macro\bfseries{\boldmath}
\makeatother

\title{From friction scaling to an efficient method for estimating bubble wall velocity}
\author[1,2]{Tomasz~Krajewski,}
\author[2,3]{Marek~Lewicki,}
\author[4]{Marco Merchand}
\author[3]{Ignacy~Nałęcz}
\author[3]{and Mateusz~Zych}

\affiliation[1]{Institute of Fundamental Technological Research of the Polish Academy of Sciences, ul. Pawi\'nskiego 5B, 02-106 Warsaw, Poland}
\affiliation[2]{Nicolaus Copernicus Astronomical Center of the  Polish Academy of Sciences, ul. Bartycka 18, 00-716 Warsaw, Poland}
\affiliation[3]{Faculty of Physics, University of Warsaw, ul.\ Pasteura 5, 02-093 Warsaw, Poland}
\affiliation[4]{Kobayashi-Maskawa Institute for the Origin of Particles and the Universe,\\
Nagoya University, Tokai National Higher Education and Research System,  \\
Furo-cho Chikusa-ku, Nagoya, 464-8602 Japan}

\emailAdd{tkrajews@ippt.pan.pl}
\emailAdd{marek.lewicki@fuw.edu.pl}
\emailAdd{merchand.medina.marco.antonio.n1@f.mail.nagoya-u.ac.jp}
\emailAdd{ignacy.nalecz@fuw.edu.pl}
\emailAdd{mateusz.zych@fuw.edu.pl}

\abstract{
We present a unified description of first-order cosmological phase transition dynamics that links the phenomenological friction model employed in hydrodynamic simulations to the microscopic treatment based on Boltzmann equations. We derive an approximate analytical expression for the chemical potential and demonstrate that the resulting friction parameter $\tilde{\eta}$ follows a simple power-law dependence on the transition strength ($\propto v_n^4/T_n^4$). Incorporating this scaling into a phenomenological framework accurately reproduces the terminal wall velocities obtained from the full microscopic analysis performed using \texttt{WallGo}~\cite{Ekstedt:2024fyq}. This approach offers an efficient method to quantify out-of-equilibrium contributions to friction and reliably estimate bubble-wall velocities.
}
\begin{document}
\maketitle
\section{Introduction}
Cosmological first-order phase transitions constitute an important class of phenomena that may have occurred in the early Universe. Investigations of such events provide a natural bridge between high-energy particle physics and cosmological observations. The transitions arise in many extensions of the Standard Model (SM) and proceed via the nucleation of bubbles of the new phase in the Universe filled with the old one. Their evolution can drive the system out of thermal equilibrium, transferring significant amounts of latent heat to the cosmic plasma and potentially giving rise to observable signatures. In particular, cosmological phase transitions can produce conditions required for generating the baryon asymmetry \cite{Kuzmin:1985mm, Cohen:1993nk, Rubakov:1996vz, Morrissey:2012db} and act as efficient sources of stochastic gravitational waves within the reach of upcoming detectors \cite{Caprini:2015zlo, Caprini:2019egz, Badurina:2021rgt, LISACosmologyWorkingGroup:2022jok, Caprini:2024ofd}. A key aspect controlling these processes is the dynamics of expanding bubbles, in particular the terminal velocity of their walls. Despite its central role, obtaining this quantity from first principles remains a non-trivial problem.

Recent studies have shown that within the local thermal equilibrium (LTE) approximation, the wall velocity in the steady state can be determined in a relatively simple manner \cite{BarrosoMancha:2020fay,Ai:2021kak, Ai:2023see}. The method relies on entropy conservation and provides an additional matching condition that constrains the thermodynamic quantities across the bubble interface. Since any non-equilibrium friction can only reduce the wall velocity, the LTE prediction, if achievable~\cite{Krajewski:2024gma}, could be interpreted as an upper bound on the result obtained from a~fully out-of-equilibrium treatment. 

This approach can be extended beyond the LTE approximation, as demonstrated in~\cite{Ai:2024btx,Krajewski:2024zxg}, where the entropy production rate is related to the resulting wall velocity. The entropy generated during the transition can be expressed in terms of the dissipative friction acting on the wall, which is typically modelled using a phenomenological ansatz \cite{Ignatius:1993qn}. This framework is widely used in modelling gravitational-wave spectra sourced by sound waves 
\cite{Hindmarsh:2013xza, Hindmarsh:2015qta, Hindmarsh:2017gnf,Cutting:2019zws, Correia:2025qif}. Nevertheless, it involves a free parameter encoding the strength of out-of-equilibrium interactions between the wall and the plasma. Since this parameter is not known a priori, it limits the predictive power of this approach in precise phenomenological studies of specific models.

On the other hand, a number of methods for microscopic determination of the out-of-equilibrium friction have been developed in recent years. One common approach is based on solving Boltzmann equations across the bubble wall, using a truncated expansion in momentum space around the equilibrium distribution, proposed by Grad in~\cite{Grad:1949zza}. In the widely used fluid approximation~\cite{Moore:1995si, Laurent:2020gpg}, the expansion is truncated at linear order, and deviations from equilibrium are parametrized in terms of a chemical potential, as well as temperature and velocity perturbations. More recently, this framework has been improved and extended to higher orders~\cite{Dorsch:2021nje, Dorsch:2023tss, Dorsch:2024jjl}, exhibiting generally good convergence.

Another class of methods relies on the spectral decomposition of solution to the Boltzmann equation in an orthogonal functional basis. While these approaches are computationally more demanding, they remain robust even when the plasma inside the wall is far from thermal equilibrium~\cite{Ai:2024btx}. They were first introduced in~\cite{Laurent:2022jrs}, where the authors found the solution to the Boltzmann equations in a basis of Chebyshev polynomials defined on a compactified spatial domain. This choice exploits the fact that out-of-equilibrium effects vanish sufficiently far away from the wall. The framework was later automated in the state-of-the-art package~\texttt{WallGo}~\cite{Ekstedt:2024fyq}. An alternative approach was proposed in~\cite{DeCurtis:2023hil, Branchina:2025adj}, where spectral decomposition of the collision operator was combined with an expansion of the coefficient eigenmodes to solve the Boltzmann equations.


The aim of this work is to establish a consistent connection between phenomenological and microscopic approaches to the dynamics of first-order cosmological phase transitions. After introducing the benchmark scalar singlet model in section \ref{sec:benchmark}, we present the hydrodynamic framework describing the propagation of expanding bubble walls in section \ref{sec:hydro}. We focus on the non-LTE limit and investigate phenomenological ans{\"a}tze for entropy production at the bubble wall in section \ref{sec:probing_ans}, providing a comparison with results obtained from the full Boltzmann treatment implemented in \texttt{WallGo} \cite{Ekstedt:2024fyq}. In the following sections \ref{sec:Frict} and \ref{sec:Frict_anal}, we derive the effective friction acting on the wall from the Boltzmann equations, introduce an approximate analytical solution for the chemical potential, and explain how the friction coefficient scales with the parameters characterizing the phase transition. This analysis leads to a simple power-law relation between the effective friction parameter and the order parameter to temperature ratio, which constitutes one of the main results of this work. Finally in section \ref{sec:num_results}, we compare different methods for determining the terminal wall velocity and demonstrate that a phenomenological description with the correct scaling of the friction term reproduces the results of the microscopic calculation with good accuracy. This provides a fast and reliable framework to estimate out-of-equilibrium corrections to the bubble-wall velocity and to classify the resulting solutions as deflagrations, hybrids, or detonations. We summarize our findings and describe the resulting method in section \ref{sec:summary}.

\section{Benchmark model}\label{sec:benchmark}

As a concrete setup for our analysis, we consider the SM extended by a real scalar singlet field, commonly referred to as the xSM~\cite{McDonald:1993ey, Espinosa:1993bs, Espinosa:2007qk, Profumo:2007wc, Espinosa:2011ax, Barger:2011vm, Cline:2012hg, Alanne:2014bra, Curtin:2014jma, Vaskonen:2016yiu, Kurup:2017dzf, Beniwal:2017eik, Beniwal:2018hyi, Lewicki:2021pgr, Niemi:2021qvp, Gould:2024jjt, Ramsey-Musolf:2024ykk}. This framework provides one of the simplest and most widely studied examples in which a first-order electroweak phase transition can occur, and has been used as a benchmark scenario in studies of bubble-wall dynamics, to which we further refer to refs.~\cite{Ekstedt:2024fyq,vandeVis:2025plm}.

The scalar sector consists of the SM Higgs doublet and an additional real scalar singlet, assumed to be invariant under a $\mathbb{Z}_2$ symmetry. At tree level, the scalar potential takes the form
\begin{equation}
\label{eq:tree-level}
    V(h,s) = \frac{1}{2} \mu_h^2 h^2 + \frac{1}{4}\lambda_h h^4 + \frac{1}{4} \lambda_{hs} h^2 s^2 + \frac{1}{2} \mu_s^2 s^2 + \frac{1}{4}\lambda_{s}s^4.
\end{equation}
The free parameters of the scalar sector are the singlet mass $m_s$, its self-coupling $\lambda_s$, and the portal coupling $\lambda_{hs}$, while the Higgs parameters are fixed to reproduce the observed mass and vacuum expectation value.

Thermal effects are incorporated through the finite-temperature effective potential $V_{\mathrm{eff}}(h,s,T)$, which in the present analysis is obtained by supplementing the tree-level potential with temperature-dependent mass corrections. In practice, this amounts to replacing the zero-temperature mass parameters by
\begin{equation}
\mu_h^2(T) \coloneqq \mu_h^2 + c_h^2 T^2 \qquad \textrm{and} \qquad \mu_s^2(T) \coloneqq \mu_s^2 + c_s^2 T^2.
\end{equation}
The corresponding coefficients are given by
\begin{equation}
c_h^2 = \frac{1}{48}\left(9g^2 + 3g'^2 + 12y_t^2 + 24\lambda_{h} + 2 \lambda_{hs}\right),
\qquad
c_s^2 = \frac{1}{12} \left( 2 \lambda_{hs} + 3\lambda_s \right),
\end{equation}
where $g$ and $g'$ denote the electroweak gauge couplings and $y_t$ is the top-quark Yukawa coupling. A detailed discussion of the validity and limitations of this approximation can be found in ref.~\cite{Espinosa:2011ax}.

The phase transition proceeds through thermal tunneling, and the nucleation temperature $T_n$ is defined by the standard criterion $S_3(T_n)/T_n \simeq 140$~\cite{Caprini:2019egz}, where $S_3$ denotes the three-dimensional Euclidean action evaluated on the corresponding bounce solution.  In the numerical analysis, we adopt the same benchmark points as in ref.~\cite{Krajewski:2024gma}, to which we refer for further details on the construction of the effective potential and the analysis of the phase transition.

Following refs.~\cite{Cline:2021iff,Laurent:2022jrs, Krajewski:2024zxg}, we assume that non-equilibrium friction arises solely from the interaction between the cosmic plasma and the Higgs field. Consequently, in the following we identify the background field relevant for the bubble-wall dynamics with the Higgs direction and set $ h=\phi$.

\section{Hydrodynamics of bubble expansion}\label{sec:hydro}

In order to describe the evolution of the bubbles, one has to introduce a dynamical model for the scalar field and the primordial plasma filling the Universe. The primordial plasma is modeled as a relativistic fluid, with stress–energy tensor 
\begin{equation}\label{eq:T_fl}
    T^{\mu \nu}_{\textrm{fluid}} = w u^{\mu}u^{\nu} - g^{\mu\nu}p.
\end{equation}
Here, $u^\mu$ is the plasma four-velocity, while $p$ and $w$ denote the pressure and enthalpy density respectively. The equilibrium values of these quantities are obtained from the effective potential
\begin{align}
    p&=-V_{\rm eff}(\phi,\,T),&w&=-T\frac{dp}{dT}, &s&=\frac{dp}{dT},
\end{align}
where we also defined the local entropy density $s$ for later use.

The stress–energy tensor of the scalar field is given by
\begin{equation}
T_{\phi}^{\mu\nu}=\partial^\mu\phi\partial^\nu\phi-g^{\mu\nu}\frac{1}{2}\partial_\alpha \phi \partial^\alpha \phi,
\end{equation}
where $g_{\mu\nu}$ is the metric tensor. The associated equation of motion reads
\begin{equation}
\partial_\mu T_{\phi}^{\mu\nu}=\frac{\partial V_{\rm eff}}{\partial \phi}\partial^\nu \phi+\mathcal{F}\,\partial^\nu \phi,
\end{equation}
with $\mathcal{F}$ denoting the out-of-equilibrium friction term that slows down the bubble wall.

The primordial plasma and the scalar field are coupled through out-of-equilibrium friction and indirectly through thermal interactions. Away from the phase boundary, the scalar field is approximately constant, and the equations of motion of the plasma reduce to the relativistic Euler equations. These equations can be solved to obtain stationary profiles of the plasma shock waves induced by the expanding bubble walls. Assuming spherical symmetry and exploiting the absence of any characteristic scale in the problem, the Euler equations reduce to a system of two ordinary differential equations for the plasma velocity $v$ and the enthalpy density $w$,
\begin{equation}\label{eq:v_diff}
    \begin{aligned}
        &2\frac{v}{\xi}=\gamma^2(1-v\xi)\Big[\frac{\mu^2(\xi, v)}{c^2}-1 \Big]\partial_\xi v,\\
        &\partial_\xi w=w\Big(1+\frac{1}{c^2}\Big)\gamma^2\mu \partial_\xi v.
    \end{aligned}
\end{equation}
Here the self-similar variable $\xi \equiv r/t$ is defined as the distance from the bubble centre divided by the time elapsed since nucleation, and $c$ denotes the speed of sound. The plasma velocity $v$ is defined in the rest frame of the bubble center, and
\begin{align}
    \gamma &\equiv \frac{1}{\sqrt{1 - v^2}},
    &
    \mu(\xi, v) &\equiv \frac{\xi - v}{1 - \xi v}.
\end{align}

The system \eqref{eq:v_diff} can be solved numerically%
\footnote{The boundary conditions follow from the requirement that the primordial plasma remains unperturbed far away from the bubble center.} %
in the symmetric and broken phases, provided that appropriate matching conditions for the thermodynamic plasma profiles are imposed. Two of these conditions follow directly from energy and momentum conservation across the bubble wall,
\begin{align}\label{eq:matching}
w_{+}\gamma_{+}^2 v_{+} &= w_{-}\gamma_{-}^2 v_{-},\\
w_{+}\gamma_{+}^2 v_{+}^2 + p_{+} &= w_{-}\gamma_{-}^2 v_{-}^2 + p_{-},
\end{align}
with the subscripts $+$ and $-$ denoting quantities evaluated just in front of and just behind the phase boundary, respectively. The third condition required to match the solutions across the bubble wall can be obtained by considering the entropy balance across the wall \cite{Ai:2024btx,Krajewski:2024zxg}
\begin{equation}\label{eq:match_DSb}
    \frac{T_{+}}{T_{-}}=\frac{\gamma_{-}}{\gamma_{+}}\,\left(1+\tilde{\eta}\,\gamma_{+}\,v_{+}\right),
\end{equation}
with
\begin{equation}
    \tilde{\eta}\equiv\frac{T_{+}\Delta S}{w_{+}\gamma_{+}^2v_{+}^2},
\end{equation}
where $\Delta S$ denotes the entropy produced within the wall per unit area of the wall surface.

If the dissipative processes that slow down the bubble wall are localized, $\Delta S$ can be written as an integral of the local entropy source across the bubble wall
\begin{equation}\label{eq:Div_s}
    \Delta S\equiv\int_{\rm wall}dz\,\partial_\mu (s u^\mu).
\end{equation}
The entropy source can be, in general, a complicated function of position and thermodynamic variables
\begin{equation}
    \partial_\mu (s u^\mu)\equiv f_s.
\end{equation}
However, if there is no accumulation of energy and momentum in out-of-equilibrium effects%
\footnote{Here we neglect out-of-equilibrium plasma excitations (not to be confused with $\mathcal{F}$), assuming that $\partial_\mu T^{\mu\nu}_{\textrm{fluid}} = -\partial_\mu T^{\mu\nu}_{\phi}$. In principle, these effects should be included in a more complete treatment, potentially complicating the relation between the out-of-equilibrium friction and the entropy source. However, as will become clear in sec.~\ref{sec:probing_ans}, our simplified approach is sufficient to predict the final shape of the shock waves with good accuracy.}
$f_s$ can be obtained directly from the out-of-equilibrium friction acting on the scalar field (for the proof, see appendix A of ref.~\cite{Ai:2023see})
\begin{equation}\label{eq:frict_to_s}
f_s = \frac{1}{T}(u^\mu \partial_\mu \phi)\times\mathcal{F}.
\end{equation}
Thus, the problem of determining the shock-wave profiles and the wall velocity $v_w$ reduces to finding an explicit expression for $\mathcal{F}$.

\section{Probing phenomenological friction}\label{sec:probing_ans}

Lorentz symmetry tightly constrains the possible form of local, near-equilibrium dissipation in relativistic field–fluid systems. The lowest-order effective friction term in $\partial_\mu \phi$ commonly used in hydrodynamic simulations reads~\cite{Hindmarsh:2013xza,Hindmarsh:2015qta,Hindmarsh:2017gnf,Cutting:2019zws,Cutting:2022zgd,Krajewski:2023clt}
\begin{equation}\label{eq:F_ansatz}
    \mathcal{F}_{\rm (ans)}=\eta\,u^\mu\partial_\mu\phi,
\end{equation}
where $\eta$ is a phenomenological parameter encoding the strength of friction and is, in general, a function of $T$ and $\phi$. The corresponding expression for the entropy source can be obtained from eq.~\eqref{eq:frict_to_s} and reads
\begin{equation}\label{eq:s_ansatz}
    f_s^{\rm (ans.)}=\frac{\eta}{T}\,(u^\mu\partial_\mu\phi)^2.
\end{equation}
In the literature, the coefficient $\eta$ is commonly assumed to be constant~\cite{Ignatius:1993qn,Kurki-Suonio:1995yaf,Balaji:2020yrx}. Under this assumption, the entropy source is dominated by gradients of the scalar field, and deviations from LTE remain small. Although these assumptions are difficult to justify from first principles, the resulting ansatz has proven remarkably successful in reproducing stationary bubble-wall solutions in hydrodynamic treatments~\cite{Krajewski:2023clt}.

In principle, the out-of-equilibrium friction $\mathcal{F}$ between the scalar field and the plasma, as well as the associated entropy source, can be derived directly from microphysics. This idea was pursued in the recent work~\cite{Ekstedt:2025awx}, where the Chapman--Enskog (CE) expansion of the Boltzmann equations was used to motivate a local, near-equilibrium expression 
\begin{equation}\label{eq:F_Ch_En}
    \mathcal{F}_{\rm (CE)}=\kappa\,\phi^2\,(u^\mu\partial_\mu\phi).
\end{equation}
where $\kappa\,T$ is a constant parameter fixed by the particle content of the microscopic model. In this limit, the entropy production rate reads
\begin{equation}\label{eq:s_Ch_En}
    f_s^{\rm (CE)}=\frac\kappa T\phi^2\,(u^\mu\partial_\mu\phi)^2.
\end{equation}
While the entropy source $f_s^{\rm(ans.)}$, inferred from the phenomenological friction $\mathcal{F}_{\rm (ans)}$ does not coincide with $f_s^{\rm (CE)}$, the resulting dissipation exhibits a similar local structure, controlled by the same power of the scalar field gradient.


\begin{figure}[t]
    \centering
    \includegraphics[width=0.45\linewidth]{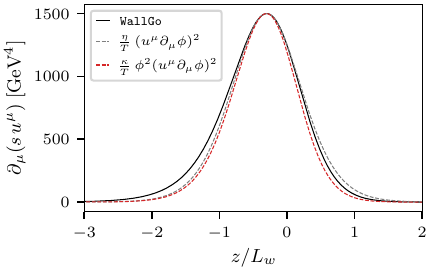}\hspace{4mm}
    \includegraphics[width=0.45\linewidth]{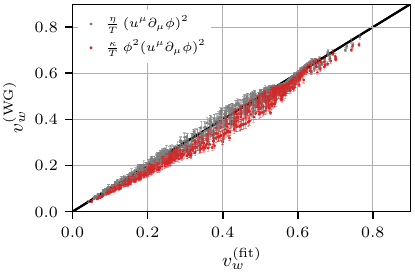}
    \caption{\textit{
    \textbf{Left panel:} Comparison of entropy production sources for the xSM benchmark obtained from the numerical solution of the Boltzmann equations, using \texttt{WallGo}~\cite{Ekstedt:2024fyq}, with two approximate descriptions: the phenomenological ansatz \eqref{eq:s_ansatz} and the CE expansion \eqref{eq:s_Ch_En}. In both approximate cases, the overall normalization of the entropy source is fitted to the numerical Boltzmann result. This illustrates that both approaches accurately reproduce the shape of the microscopic entropy source. \\
    \textbf{Right panel:} Terminal bubble-wall velocities computed using the hydrodynamic matching conditions with entropy production given with eq.\,\eqref{eq:match_DSb}, where the friction parameters are fixed by the fits shown in the left panel, compared to the wall velocities obtained directly from \texttt{WallGo}. 
}}
    \label{fig:ans_vs_WG}
\end{figure}

In order to critically assess the validity of the phenomenological ansatz \eqref{eq:s_ansatz} and the profile obtained from CE expansion \eqref{eq:s_Ch_En}, we confront them with the friction profile obtained from a fully microscopic treatment of out-of-equilibrium dynamics. To this end, we employ the state-of-the-art Boltzmann solver \texttt{WallGo}~\cite{Ekstedt:2024fyq}. The cross-check is held solely for solutions on the deflagration/hybrid branch, for which numerical uncertainties are better controlled. Nevertheless, both phenomenological friction and \texttt{WallGo} can also be used to extract wall velocity of detonation solutions, albeit with lower accuracy~\cite{Ekstedt:2024fyq,Krajewski:2024zxg}. 

For a given benchmark point in the xSM parameter space, \texttt{WallGo}~\cite{Ekstedt:2024fyq} provides both the stationary field–fluid configuration and the corresponding microscopic entropy production profile $f_s$ (or equivalently, the friction)  across the wall. We then adjust the parameters $\eta$  and $\kappa$ in the phenomenological ansatz \eqref{eq:s_ansatz} and in the approximate local expression \eqref{eq:s_Ch_En} to best match the numerical friction profile. The resulting comparison between the microscopic friction and the fitted profiles is shown in the left panel of fig.~\ref{fig:ans_vs_WG}. 
As seen in the figure, the three curves are in very good agreement. This behaviour can be traced back to the dominance of the scalar field gradient in the source term of the Boltzmann equation. Consequently, it is expected that any friction ansatz proportional to $u^{\mu}\partial_{\mu}\phi$ provides a good approximation. In Section \ref{sec:Frict_anal}, we demonstrate that this proportionality naturally emerges from the kinetic theory framework.


Using the fitted values of $\eta$ and $\kappa$, one can recompute the wall velocity $v_w$. Technically, this is done by evaluating the integral defining $\Delta S$ for the entropy sources $f^{\rm (ans.)}_s$ and $f^{\rm (CE.)}_s$. In principle, evaluating this integral requires knowledge of the scalar field, temperature, and plasma velocity profiles across the bubble wall. However, for moderate and weak transitions, the temperature and plasma velocity change across the wall is mild, and one can approximate their profiles by constant values $T_+$, $v_+$ (or $T_-$, $v_-$). The scalar field profile is well approximated with the hyperbolic tangent ansatz~\cite{Moore:1995si} 
\begin{equation}
    \phi(z) = \frac{\upsilon_n}{2}\left[1-\tanh(z/L_w)\right], 
\end{equation}
where $\upsilon_n$ denotes the Higgs vacuum expectation value (VEV) at $T_n$, while $L_w$ is the profile width. Although $L_w$ in general depends on the out-of-equilibrium interactions, we found that a reasonable estimate of this parameter is obtained in the LTE limit~\cite{Huber:2013kj, Krajewski:2024zxg}
\begin{equation}\label{eq:Lw}
    L_w=\frac{\upsilon_n}{\sqrt{8 V_b}},
\end{equation}
with $V_b$ being the potential barrier height at the critical temperature. Once an expression for $\Delta S$ is obtained, it is used to relate $\eta$ and $\kappa$ to $\tilde\eta$
\begin{equation}\label{eq:eta_tilde}
    \tilde\eta\approx\begin{cases} \eta\times\frac{ \upsilon_n^2}{3\, w_{s} L_w}\qquad&\text{for}\quad f_s^{\rm (ans.)} \\[1mm]
    \kappa\times\frac{ \upsilon_n^4}{10\, w_{s} L_w}\quad&\text{for}\quad f_s^{\rm (CE)}\,.
    \end{cases}
\end{equation}
Here, we approximate the enthalpy density and temperature just in front of the wall, $w_+$ and $T_+$, by their asymptotic values in the symmetric phase, $\omega_s$ and $T_n$. 

Once $\tilde\eta$ is computed, we determine the terminal wall velocity $v_w$ by solving the system of equations in eq.~\eqref{eq:v_diff} using the numerical algorithm developed in our previous work~\cite{Krajewski:2024zxg}. The implementation of this method is publicly available in the form of an open-source code~\cite{Repo:2024in}, which performs the hydrodynamic matching and determines $v_w$ for a given value of the effective friction parameter.

The plot summarizing our computation for a random sample of xSM benchmarks is shown in the right panel of fig.~\ref{fig:ans_vs_WG}. At each point the $v_w$ computed based on entropy sources $f_s^{\rm (ans.)}$ and $f_s^{\rm (CE)}$ is compared to the wall velocity obtained directly from \texttt{WallGo} \cite{Ekstedt:2024fyq}. We observe that our simple scheme allows for precise estimate of the wall velocity regardless of the entropy source details. This should not come as a surprise since both CE and standard ansatz profiles approximately reproduce the shape of the original numerical entropy source, as shown in the left panel of fig. \ref{fig:ans_vs_WG}.

\begin{figure}[t]
    \centering
    \includegraphics[width=0.85\linewidth]{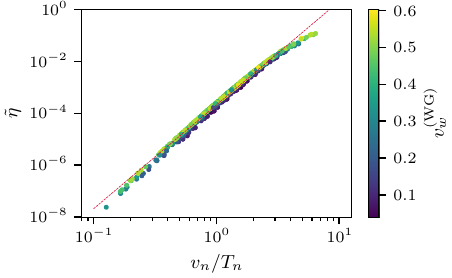}
    \caption{\textit{The rescaled friction parameter $\tilde\eta$, extracted from the \texttt{WallGo}~\cite{Ekstedt:2024fyq} friction, shown as a function of the ratio $\tfrac{v_n}{T_n}$. The red dashed line indicates the power-law relation given in eq.~\eqref{eq:power_low}, with $\zeta^{\rm (WG)} = 2.1\times10^{-4}$ obtained from fitting to the data.}}
    \label{fig:tilde_v_T}
\end{figure}

Motivated by this agreement, we now turn to a more detailed analysis of the effective friction itself. In particular, we investigate how the rescaled friction coefficient $\tilde\eta$, extracted from \texttt{WallGo}~\cite{Ekstedt:2024fyq}, depends on the ratio of the Higgs vacuum expectation value at nucleation to the nucleation temperature. As shown in fig.~\ref{fig:tilde_v_T}, we find that this quantity exhibits a remarkably simple power-law behavior over a wide range of benchmark points, given by
\begin{equation}\label{eq:power_low}
\tilde\eta = \zeta \left(\frac{v_n}{T_n}\right)^4,
\end{equation}
with $\zeta^{\rm (WG)} = 2.1 \times 10^{-4}$. %

In section \ref{sec:Frict_anal}, we show that this empirical scaling can be understood from a microscopic perspective. To this end, we analyse the Boltzmann equations governing the out-of-equilibrium dynamics of the plasma and demonstrate that the power-law behaviour in eq.~\eqref{eq:power_low} naturally emerges from the underlying kinetic theory. Moreover, we present an independent determination of the coefficient $\zeta$, which allows for an accurate and computationally efficient estimate of the terminal wall velocity without the need to solve the full set of Boltzmann equations.



\section{Microscopic friction from kinetic theory}\label{sec:Frict}

To describe the dynamics of the growing bubble, it is essential to evaluate the total force exerted on its wall by the surrounding plasma. At the microscopic level, this force arises from interactions between the scalar field driving the transition and the particles whose masses depend on its local value. The resulting equation of motion for the scalar field in the presence of a relativistic plasma can be written as
\begin{equation}
\Box \phi + \frac{\partial V_{\textrm{eff}}}{\partial \phi} + \sum_{i} n_i \frac{\textrm{d} m_i^2(\phi)}{\textrm{d}\phi}\int\frac{\textrm{d}^3 k}{(2\pi)^3 2 E_i}\delta f_i(k,x) = 0,
\label{eqn1}
\end{equation}
where $\delta f_i$ denotes the  out-of-equilibrium part of the particle distribution function, and the index $i$ runs over all the particle species interacting with $\phi$. The last term on the left-hand side encodes the friction exerted by the perturbation of the plasma particles. In principle, all particle species whose masses vary across the wall are pushed out of equilibrium as they traverse it. Here, however, following~\cite{Cline:2020jre,Ekstedt:2024fyq,Laurent:2020gpg, Laurent:2022jrs} we only focus on the species that yield the leading frictional effects and assume that the remaining plasma components stay close to equilibrium.
Under this assumption, the out-of-equilibrium distribution functions of the relevant species can be obtained by solving the Boltzmann equation. For a single particle species, it takes the form
\begin{equation}
\label{eq:bolzmann}
\left[p^\mu\partial_\mu-\frac12(\partial_\mu m^2)\partial_{p_\mu}\right]f(x^\nu, p^\nu)=-\mathcal{C}[f]
\end{equation}
where $m^2$ denotes the squared mass of the considered (typically heavy) particle species, while $\mathcal{C}[f]$ is the collision term describing interactions\footnote{Here we only consider $2\to2$ processes.  This choice follows the standard approximation adopted in most existing micro-physical calculations of bubble–wall friction, which also neglect explicit 
$1\rightarrow2$ processes. A fully systematic leading‑order treatment in hot gauge theories would in general require incorporating effective collinear $1\rightarrow2$ processes together within 
$2\rightarrow2$ scatterings. Assessing the quantitative impact of these additional channels on electroweak bubble dynamics lies beyond the scope of the present work.
} with the surrounding plasma, given by
\begin{equation}
    {\cal C}[f]=\frac{1}{4N_p}\sum_{j}\int d\Omega_{ k}\,d\Omega_{ p'}\,d\Omega_{ k'}|{\cal M}_j|^2 (2\pi)^4\delta^{(4)}(p + k - p' -k'){\cal P}[f]\,.
\end{equation}
Here $N_p$ denotes the number of particle degrees of freedom, the sum runs over all the relevant processes in the plasma, and $\mathcal{M}_j$ is the spin-averaged matrix element for a given process. We have introduced the shorthand notation
\begin{equation*}
    d\Omega_{k}\equiv\frac{d^3{\bf k}}{(2\pi)^3 2 E_k}, \quad\text{and}\quad  
    \mathcal{P} = f_p f_k(1\pm f_{k'})(1\pm f_{p'})-f_{p'} f_{k'}(1\pm f_{k})(1\pm f_{p}),
\end{equation*}
where the statistical factors $\pm$ account for Bose-Einstein and Fermi-Dirac distributions, respectively.

From now on, we restrict ourselves to the top quark, which, owing to its large Yukawa coupling, typically provides the dominant contribution to the friction acting on the bubble wall. In this case, the friction term can be written as
\begin{equation}\label{eq:frict_def}
    \mathcal{F}_{t} \equiv N_{t} \frac{dm_{t}^2}{dh}\int \frac{d^3k}{(2\pi)^3 2E}\delta f,
\end{equation}
with $N_{t}=12$ and $m_{t}= y_t h(z)/\sqrt{2}$.  The out-of-equilibrium part of the distribution function can be expanded in momenta
\begin{equation}\label{eq:fluid_ansatz}
    \delta f=(f_0')\times\left[\mu+T\,(\,\delta u\, u^\mu+\beta\,\delta \tau\, \bar{u}^\mu)p_\mu+T^2\omega^{\mu\nu}p_\mu p_\nu+\ldots\right],
\end{equation}
where the series truncated at $p^\mu$ yields the well-known fluid ansatz~\cite{Moore:1995si}. The zeroth order term in this expansion admits a simple physical interpretation as an effective chemical potential. Since it appears as the leading term in the momentum expansion, it is expected to provide the dominant contribution to the friction. Physically, this term controls the excess population of particles inside the bubble wall relative to equilibrium, which constitutes the primary source of the out-of-equilibrium force acting on the wall (see fig. 1 in ref. \cite{Laurent:2020gpg}). Motivated by this observation, we truncate the series expansion \eqref{eq:fluid_ansatz} at zero order which leads to a friction term of the form
\begin{equation}\label{eq:frict_approx}
   \mathcal{F}_{t}\approx\mathcal{F}_t^{(\mu)}  \equiv \frac{dm_{t}^2}{dh}\frac{N_{t}T^2}{2}  C_0^{1,0}(z)\mu_t(z),
\end{equation}
and considerably simplifies the Boltzmann equation \eqref{eq:bolzmann} to 
\begin{equation}\label{eq:mu}
    \gamma_w v_w  C_0^{0,0}(z)\frac{ d \mu_t(z)}{d z} + \Gamma_{0,0} \ T \mu_t(z) = \gamma_{w} v_{w} \frac{(m^2)'}{2 T^2}C_0^{1,0}(z). \footnote{In writing \eqref{eq:mu}  we followed the approximations introduced in \cite{Cline:2020jre}. In the wall frame the fluid's velocity $v$ is set by the wall $v_w$ with respect to the bubble's center.  We also neglected terms  $-T^{-2} C^{0,0}_0 \partial_z T - T^{-1} \gamma^2 C^{1, 1}_0 \partial_z v$ proportional to derivatives $\partial_z T$, $\partial_z v$ of thermodynamic variables, present on the right hand side of the equation.}
\end{equation}

Here the coefficients
\begin{equation}
    C_{0}^{0,\,0} = T^{-3}\int\frac{d^3 p}{(2\pi)^3}(-f'_0), \qquad C_{0}^{1,\,0} = T^{-2}\int\frac{d^3 p}{(2\pi)^3}\frac{1}{E}(-f'_0) ,
\end{equation}
encode the momentum integrals over the boosted Fermi-Dirac distribution function $f_v$, evaluated in the plasma rest frame \cite{Laurent:2020gpg}.
The parameter $\Gamma_{0,0}$ represents the effective interaction rate controlling the relaxation of the chemical potential toward equilibrium, whose value is fixed by the integral~\cite{Guiggiani:2024owe}
\begin{equation}
        \Gamma_{0,0}=\beta^4\frac{1}{2N_p}\int d\Omega_p\, d\Omega_k \,d\Omega_{p'}\, d\Omega_{k'}\,|{\cal M}_1|^2(2\pi)^4\delta^4(p+k-p'-k')\,\bar{\cal P}_{0}, 
\end{equation}
which follows from taking the first moment of the Boltzmann equation, namely $\beta^4\int d^3 p$. In the equation above, $\bar{\cal P}_{0}$ denotes linearized population factor
\begin{equation*}
    \bar{\cal P}_0 = f_0(\beta E_{\bar p})f_0(\beta E_{\bar k})(1\pm f_0(\beta E_{\bar p'}))(1\pm f_0(\beta E_{\bar k'}))\,,
\end{equation*}
and $|\mathcal{M}_1|^2$ stands for the matrix element of the top annihilation $t\bar t\to g\bar g$, which in the leading logarithm approximation reads
\begin{equation}\label{eq:M_scatter}
    |\mathcal{M}_1|^2=\frac{128}{3}g_s^4\left[\frac{t\,u}{(m_t^2-t)^2}+\frac{t\,u}{(m_t^2-u)^2}\right],
\end{equation}
with $g_s$ being the strong gauge coupling and $s$, $t$ denoting Mandelstam variables.

Depending on the literature source, one finds substantially different numerical values for this integral:
$\Gamma_{0,0}^{\rm (G.)}=0.00128$~\cite{Guiggiani:2024owe},
$\Gamma_{0,0}^{\rm (L.C.)}=0.00196$~\cite{Laurent:2020gpg},
$\Gamma_{0,0}^{\rm (D.H.K.)}=0.0044$~\cite{Dorsch:2021ubz}, and
$\Gamma_{0,0}^{\rm (B.M.)}=0.00899$~\cite{Moore:1995si}.
The last value is known to omit the symmetry factor in the top-annihilation diagram, thereby
overestimating $\Gamma_{0,0}$ by a factor of two~\cite{Arnold:2000dr}.
Furthermore, the results in refs.~\cite{Moore:1995si,Dorsch:2021ubz}, rely on several simplifying assumptions which enable the analytical evaluation $\Gamma_{0,0}$, but result in a less accurate value compared to the rigorous, numerical treatment in refs.~\cite{Laurent:2020gpg,Guiggiani:2024owe}%
.\footnote{The resulting overestimate of the collision matrix terms is most probably the cause
of the discrepancy between the fluid ansatz and \texttt{WallGo}~\cite{Ekstedt:2024fyq} results observed in
ref.~\cite{Ekstedt:2024fyq}. In section~\ref{sec:num_results} we introduce a simplified
version of the fluid ansatz valid for slowly moving walls, and show that it aligns reasonably
well with the results obtained with \texttt{WallGo}.} 
%
Finally, we found that the scattering-matrix definition used in~\cite{Laurent:2020gpg}
differs from \eqref{eq:M_scatter} (also used in~\cite{Guiggiani:2024owe}) by terms that are
negligible in the leading-logarithmic approximation.
Nevertheless, these subleading contributions account for the numerical discrepancy between
$\Gamma_{0,0}^{\rm (L.C.)}$ and $\Gamma_{0,0}^{\rm (G.)}$.~\footnote{We acknowledge Daniel Pinto for pointing out this difference.}
Although determining which of the two values is more accurate would require going beyond the
leading-log approximation, we choose~\cite{Guiggiani:2024owe} because it utilizes an expression
for $|\mathcal{M}_1|^2$ that is consistent with the matrix element used by \texttt{WallGo} to
compute collision terms in a Chebyshev basis~\cite{Ekstedt:2024fyq}.

\section{Analytic approximation for the chemical potential }\label{sec:Frict_anal}%


In this section we derive an analytic approximation for the  friction term in eq. \eqref{eq:frict_approx}. Motivated by the results shown in figure \ref{fig:tilde_v_T} we will then use the result to determine friction scaling with $v_n/T_n$ ratio. %

We thus begin by considering the Boltzmann equation \eqref{eq:mu} in the simplified form:
\begin{equation}
    \frac{d \mu_t(z)}{dz} + b \frac{T}{\gamma_w v_w} \mu_t(z) = \frac{c}{2 T^2} \phi(z) \phi'(z),
    \label{eq:chemicalpotentialEOM}
\end{equation} 
where $b \equiv \frac{\Gamma_{0,0}}{C_0^{0,0}}$, $c  \equiv \frac{C_0^{1,0}}{C_0^{0,0}}$ and the top quark mass is $m_t = y_t \phi(z)/\sqrt{2}$.

Assuming the hyperbolic tangent for the Higgs profile $\phi(z)$ and setting the coefficients $b \approx 0.016$, $c \approx 0.375$
to constant values,  where we approximated $C_0^{0,0}\approx 0.08$, $C_0^{1,0}\approx 0.03$ by their average over the $z$ coordinate in the range $-L_w$ to $L_w$.\footnote{We also found that $C^{1,0}\approx0.03$ at the friction peak for most benchmark points. Notice that in the high temperature limit, one obtains   $C_0^{0,0}\approx 0.083$, $C_0^{1,0}\approx 0.035$ which is close to the quoted values; however, we prefer to use their average as they vary across the bubble wall.} This simplification allows us to obtain an analytic solution
\begin{equation}
\mu_{t}(y) = c \frac{v_n^2}{T^2} \frac{1}{4} \left[ -\frac{1 + \Delta + (2 + \Delta) y}{(1 + y)^2} + (1 + \Delta) \, {}_2F_1(1, \Delta; 1 + \Delta; -y) \right],
\label{eq:mutop}
\end{equation}
where we defined $\Delta\equiv b L_w T / (2 v \gamma)$ with $y = e^{2z/L_w}$  and ${}_2F_1$ is the hypergeometric function.


Physically, $\Delta $ represents the ratio of scales of the wall width to the inverse collision strength, viz. $\Delta \sim L_w/(\Gamma^{-1})$. Thus, the limit $\Delta \gg1$ indicates that the inverse collision strength (and the mean free path, since it follows a similar scaling) greatly falls behind the Lorentz contracted bubble wall width which is the characteristic scale in the problem.  In this limit,  the derivative term in the transport equation can be neglected and we get

\begin{equation}
    \mu_{t}(z)\big|_{\Delta \gg1} =  \frac{c}{2b}\frac{\gamma_w v_w}{T^3} \phi(z)\phi'(z) + \mathcal{O}\left(\frac{1}{\Delta^2}\right),
\end{equation}
which, via eq. \eqref{eq:frict_approx}, produces a friction 
\begin{equation}\label{eq:approximate_Ch_En}
    \mathcal{F}_t \approx \frac{3}{T}\frac{(C_0^{1,0})^2}{\Gamma_{0,0} }\phi(z)^2u^{\mu}\partial_{\mu} \phi(z)\xrightarrow[]{ C^{1,0}_0\, \approx \,0.03}\frac{2.1}{T}\phi(z)^2u^{\mu}\partial_{\mu} \phi(z), 
\end{equation}
This functional form maps directly onto the phenomenological term \eqref{eq:F_Ch_En}, immediately suggesting that $\kappa =2.1/T$, and thus transparently bridging microscopic and macroscopic descriptions. Furthermore, it 
recovers the friction coefficient structure from ref.~\cite{Ekstedt:2025awx} (see their appendix B). 

Nevertheless, a closer look at eq.  \eqref{eq:approximate_Ch_En} above reveals that the approximation  $\kappa = const/T$ (across the wall) used in \cite{Ekstedt:2025awx} needs to fail at least in some cases. This is because the top quark is massive inside the bubble and the high temperature approximation is not valid there. We notice that the mass dependent factor $(C_0^{1,0})^2$ suppresses the friction inside the bubble where the massive top has mass comparable to the nucleation temperature and goes as $C_0^{1,0} \sim 2^{-5/2} \pi^{-3/2} \sqrt{\frac{m_t}{T}} e^{-m_t / T} \xrightarrow[]{m_t / T \to \infty} 0$.~\footnote{One may wonder if this observation is limited to the lowest order truncation used for deriving \eqref{eq:mu}. The moments method applied to any higher (but finite) truncation of \eqref{eq:fluid_ansatz} leads to sources with multiplicative factors containing functions $C^{-m, n}$ for $-m,n \ge 0$ with the following asymptotic behaviour
\begin{equation}
    C^{-m, n} (x) \xrightarrow[]{x \to \infty} \frac{2^{3n/2 - 3/2}}{\pi \sqrt{\pi}} \frac{n!^2}{(n+2)! \frac{n}{2}!} x^{n/2+m+3/2} e^{-x}, 
\end{equation}
which suppress the friction inside the bubble.} Additionally, for fast walls the expansion in $1/\Delta$ needs to break down due to Lorentz contraction of the wall width.


We now proceed to derive the scaling we observed in eq. \eqref{eq:power_low} by extracting the value of the top quark chemical potential at its peak. For this purpose we observe that the general expression, eq. \eqref{eq:mutop}, peaks at 
\begin{equation}
z_{\rm peak} \approx \frac{L_w}{2} \log\left[ 11\log\left(1 + \frac{0.25}{\Delta}\right) + 2 \right],
\end{equation}
 which then yields 
\begin{equation}
    \mu_{t}(z_{\rm peak}) \approx 0.07 \frac{c\, v_w \gamma_w}{b L_w T} \left( \frac{v_n}{T} \right)^2,
\end{equation}
where the two equations above were obtained by a numerical fitting procedure. 

Inserting the above result into \eqref{eq:frict_approx} and matching to the phenomenological friction ansatz eq. \eqref{eq:F_ansatz} $\mathcal{F}_t=\mathcal{F}_{\rm ans}$ gives
\begin{equation}
\eta(z) \approx \frac{0.18}{\gamma_w v_w} \frac{\phi(z)}{\phi'(z)} T^2 \mu_{\rm top}(z),
\label{eq:eta}
\end{equation}
which evaluates to
\begin{equation}
\eta\big|_{z_{\rm peak}} \approx 0.018 \frac{c}{b} T \left( \frac{v_n}{T} \right)^2.
\end{equation}
Noticing that the enthalpy scales as $T^4$ and the wall width as $1/T$ we see then from eq. \eqref{eq:eta_tilde} that $\tilde{\eta} \sim (v_n/T)^4$.

\section{An efficient method to compute bubble-wall velocity}\label{sec:num_results}


With an approximate expression for the out-of-equilibrium friction derived in section \ref{sec:Frict}, we can compute the entropy production rate inside the advancing bubble wall. The integral of this rate enters the third matching condition \eqref{eq:match_DSb}, which describes the entropy balance across the wall. Thus, finding this integral enables us to solve the system of eqs. \eqref{eq:v_diff} for the plasma thermodynamic profiles and the wall velocity $v_w$.

In practice, the equation for the chemical potential \eqref{eq:mu} which controls the friction \eqref{eq:frict_approx} and the equations for the plasma thermodynamic profiles \eqref{eq:v_diff} are solved iteratively. For the reader’s convenience, we provide below a step-by-step description of our routine:
\begin{enumerate}
\item We start from the LTE result whenever it exists. If it does not, we use the speed of sound in the Bag model, $c_s=1/\sqrt{3}$, as an initial guess for $v_w$.
\item 
We use the above input for $v_w$, to compute the chemical potential profile by solving \eqref{eq:mu} numerically, 
approximating $L_w$ with \eqref{eq:Lw} and accounting for the fact that the coefficients entering the equation vary across the wall.
\item We use the resulting chemical potential to evaluate the friction in \eqref{eq:frict_approx}. Next, we fit the ansatz \eqref{eq:F_ansatz} to \eqref{eq:frict_approx} by varying its amplitude $\eta$.
\item Finally, the fitted value of $\eta$ is converted to $\tilde{\eta}$ and used to solve the system \eqref{eq:v_diff} for $v(\xi)$ and $w(\xi)$, from which we infer $v_w$. This can be done, for example, using our publicly available script~\cite{Repo:2024in}. The resulting value of the wall velocity is then used as the starting value for the next iteration from point 2.
\end{enumerate}
The procedure is repeated until convergence to a stationary value of $v_w$ is reached. The algorithm generally converges after just a couple of iterations, yielding a unique solution to the system.%
\footnote{The solution is unique, provided we consider only deflagrations and hybrids, because for fixed latent heat $\alpha_\theta$ and enthalpy ratio $\Psi$ the wall velocity is known to be a single-valued function of $\tilde\eta$; see, e.g.,~\cite{Cutting:2021tqt,Krajewski:2023clt}.}

\begin{figure}[t]
    \centering
    \includegraphics[width=0.45\linewidth]{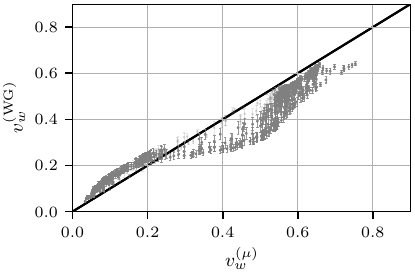}\hspace{4mm}
    \includegraphics[width=0.45\linewidth]{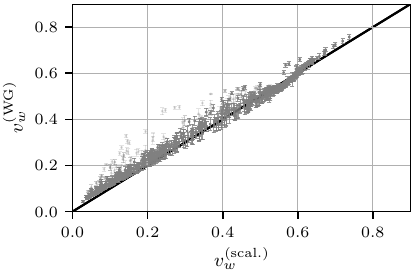}\hspace{4mm}
    \caption{\textit{\textbf{Left panel}: terminal wall velocity obtained by solving numerically the equation for the chemical potential \eqref{eq:mu} and using $v^{\rm (\mu)}_w$, versus the velocity predicted by \texttt{WallGo} code~\cite{Ekstedt:2024fyq} $v^{\rm (WG)}_w$. The only out-of-equilibrium process considered was the top annihilation $t\bar t\to g \bar g$.\\
    \textbf{Right panel}: The wall velocity computed based on the extrapolation of the phenomenological power-low rule \eqref{eq:power_low} $v^{\rm (scal.)}_w$ against \texttt{WallGo} results. The free parameter in the power-low relation was inferred from fitting to the solutions obtained from the chemical potential, and reads $\zeta^{(\mu)}\approx2.5\times 10^{-4}$. In both panels the error bars represent numerical uncertainty of the \texttt{WallGo} solutions. Benchmarks which do not satisfy the perturbativity condition $v_n/T_n>g$ were plotted with lighter shade.
    }}
    \label{fig:vw_sol_WG}
\end{figure}


Numerical results for the bubble wall velocity computed with our method and with the state-of-the-art code \texttt{WallGo}~\cite{Ekstedt:2024fyq} are shown in the left panel of fig. \ref{fig:vw_sol_WG}. For slowly moving walls with $v_w<0.3$ our method reproduces the \texttt{WallGo} results with a good accuracy. On the other hand, for larger wall velocities, the chemical potential method significantly underestimates the friction, and thus the predicted wall speed is generically higher than in \texttt{WallGo} (while still being substantially more accurate than the LTE approximation). This is not surprising, since while evaluating the friction we neglected terms proportional to powers of particle momenta, which in the bubble-wall frame become increasingly important for faster walls.

Although our approximate method is applicable only to slow walls, it can be used to determine the coefficient $\zeta$ in the scaling relation \eqref{eq:power_low}. To this end, one first evaluates $\tilde{\eta}$ for a small sample of benchmark points using the chemical potential approach and retain only the results with small wall velocities $v_w<0.3$. The coefficient $\zeta$ is then determined from this filtered sample, for which the chemical potential approximation is expected to be most accurate. The resulting value of $\zeta$ can subsequently be used to estimate $\tilde{\eta}$ and the terminal wall velocity across the full benchmark set (not just points with $v_w<0.3$). Following this procedure, we obtained
\begin{equation}
    \zeta^{(\mu)}\approx2.5\times10^{-4}.
\end{equation}
The scaling coefficient coming from this simplified procedure is similar, but not equal to the one obtained in the full analysis $\zeta^{\text{(WG)}}\approx 2.1\times10^{-4}$, see fig.~\ref{fig:tilde_v_T}.

To assess the precision of this scheme, we compared the results obtained from the scaling relation, with $\zeta$ fixed as described above, to the \texttt{WallGo}~\cite{Ekstedt:2024fyq} solutions for $v_w$, in the right panel of fig. \ref{fig:vw_sol_WG}. We found that using the phenomenological power-law scaling with the independently determined prefactor significantly improves the accuracy of our predictions relative to the results obtained solely from the chemical potential method. The best agreement is now reached for fastest solutions, but the predictions for deflagrations and hybrids with moderate $v_w$ are also noticeably improved compared to those from the bare chemical potential approach.


The procedure we propose is applicable to a wide range of SM extensions, and can be straightforwardly extended to account for many different out-of-equilibrium species. It is also extremely efficient numerically, since it only requires solving the system of linear ordinary differential equations for chemical potential(s) on a small sample of benchmarks to determine $\zeta^{(\mu)}$. Once this free parameter is fixed, $\tilde\eta$ is given analytically at every point in the model parameter space, and the computation of the wall velocity reduces to solving for eqs. \eqref{eq:v_diff} with a fixed total entropy yield, which can be easily handled e.g. with publicly available python code~\cite{Repo:2024in}.

\begin{figure}[t]
    \centering
    \includegraphics[width=0.85\linewidth]{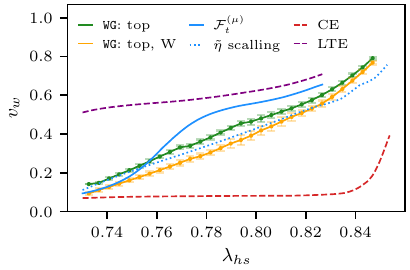}\hspace{4mm}
    \caption{\textit{Terminal wall velocity in the xSM model versus $\lambda_{HS}$ coupling for \mbox{$m_s=90$ GeV} and \mbox{$\lambda_s=1$} (c.f. with ref. \cite{vandeVis:2025plm}). 
    Dashed, purple line shows the LTE limit, where out-of-equilibrium effects are neglected. Dashed, red line corresponds to the results obtained with near equilibrium approximation for the entropy source \eqref{eq:s_Ch_En} with theoretically derived amplitude $\kappa=2.3/T_n$~as in \cite{Ekstedt:2025awx}. The solid blue line shows the approximate method based on the estimation of the chemical potential \eqref{eq:frict_approx}, assuming the only particle out of equilibrium is the top. The dotted blue line was obtained by extrapolating the phenomenological scaling of $\tilde{\eta}$ \eqref{eq:power_low}, with the coefficient inferred from chemical potential method (see discussion in the text). The orange and green lines depict full numerical solutions to the Boltzmann equations obtained with \texttt{WallGo}~\cite{Ekstedt:2024fyq}, accounting only for the out-of-equilibrium top (green line) and for out-of-equilibrium top together with massive gauge bosons (orange line). The error bars on these lines show the uncertainties of the numerical solutions.}}
    \label{fig:vw_uncertainty}
\end{figure}

To summarize our discussion, let us comment on the error associated with the approximate methods proposed in this paper. In fig. \ref{fig:vw_uncertainty} we compare the precision of different approaches for evaluating the terminal wall velocity against direct solutions of the Boltzmann equations obtained with \texttt{WallGo}~\cite{Ekstedt:2024fyq,vandeVis:2025plm}. At low velocities, all approaches yield very similar predictions. However, beyond $v_w\gtrsim 0.3$ the chemical potential method (blue solid line) becomes inaccurate. This is expected, since the characteristic momentum scale grows with $v_w$, whereas our approximate expression for out-of-equilibrium friction \eqref{eq:frict_approx} captures only the leading term in a low-momentum expansion. Nevertheless, the validity range of the phenomenological rule \eqref{eq:power_low} for the friction-parameter scaling extends over almost the entire parameter space in which deflagration and hybrid solutions exist. It is therefore the most accurate simple approximation for the wall dynamics that does not require solving the Boltzmann equations. As shown in fig. \ref{fig:vw_uncertainty}, the error associated with using \eqref{eq:power_low} is not larger than the uncertainty induced by neglecting the impact of gauge bosons on the wall dynamics (green curve). Finally, we checked that in our model the approximate local results motivated by the CE expansion~\cite{Ekstedt:2025awx} significantly overshoot the dissipative friction leading to inadequate predictions for $v_w$ (red dashed line).

\section{Summary\label{sec:summary}}
In this work, we established a quantitative connection between phenomenological and microscopic descriptions of the interaction between the scalar field and the ambient plasma during first-order cosmological phase transitions. Working within a benchmark scenario provided by the xSM model, we constructed the hydrodynamic framework that describes the expanding bubble walls and assessed commonly used phenomenological prescriptions for entropy production at the phase interface. Then these results were  systematically compared with the full Boltzmann treatment implemented in the \texttt{WallGo} package~\cite{Ekstedt:2024fyq}, allowing us to evaluate the accuracy of simplified approaches commonly employed in phenomenological studies. 

Our analysis shows that such simplified descriptions provide a remarkably accurate and computationally efficient alternative to the full kinetic treatment. In particular, we find that the effective friction parameter $\tilde{\eta}$ governing the wall dynamics exhibits a simple power-law scaling with the order-parameter-to-temperature ratio at nucleation
\begin{equation}
    \tilde{\eta} = \zeta\, \left(\frac{\upsilon_n}{T_n}\right)^4, \qquad\text{with}\quad \zeta^{\rm(WG)}=2.1\times10^{-4}.
\end{equation}

To understand the origin of this relation, we developed an analytical approximation to the Boltzmann equation based on the chemical potential of the plasma species. Within this framework, the relevant out-of-equilibrium effects can be captured without solving the full set of Boltzmann equations. Remarkably, this approach reproduces the observed scaling behaviour and yields an independent estimate of the coefficient
$
\zeta^{(\mu)}=2.5\times10^{-4}, 
$
demonstrating that the dominant transport effects responsible for friction can be reliably captured within this simplified description.

Although our analysis has been carried out within the xSM extension, the methodology developed in this work is not tied to this specific setup. In particular, we explain how the procedure for extracting the effective friction parameter from microscopic plasma dynamics can be generalized to other particle-physics scenarios featuring first-order phase transitions and similar bubble-wall propagation dynamics.

Finally, we applied our findings to the determination of the terminal bubble-wall velocity. In this context, we compared several methods for computing the wall velocity and demonstrated that a phenomenological description incorporating the extracted friction scaling reproduces the results of the full microscopic calculation with high accuracy. Consequently, the framework developed in this work provides a fast and reliable approach for estimating non-equilibrium corrections to bubble-wall velocities in first-order cosmological phase transitions, enabling efficient exploration of particle-physics models relevant for early-Universe cosmology.

\begin{acknowledgments}
We thank Benoit Laurent for fruitful correspondence and for his valuable assistance with the \texttt{WallGo} package, and Daniel Pinto for discussions on the subtleties of the transport equations and collision term moments. We also acknowledge Philipp Schicho for useful discussions during his visit to Warsaw.

ML, MZ and IN were supported by the Polish National Science Centre project number \linebreak[4]\mbox{2023/50/E/ST2/00177}. The work of ML was supported by TEAMING grant Astrocent Plus (GA: 101137080) funded by the European Union. MM is funded by KMI startup fund and by KMI/FlaP Young Researchers Grant. IN was supported by the Polish National Science Centre research grants no. 2020/38/E/ST2/00243. TK was supported by Polish National Science Centre grants 2019/33/B/ST9/01564 and 2023/51/B/ST9/00943. MZ and IN were also supported with the IDUB Early Universe scholarships.
\end{acknowledgments}

\bibliographystyle{JHEP}
\bibliography{References} 
\end{document}